\documentclass{aa}
\usepackage{graphicx}
\usepackage{txfonts}
\begin{document}
\title{On the VLBI measurement of the Solar System acceleration}
\author{O. Titov\inst{1} \and S. Lambert\inst{2} }
\institute{Geoscience Australia, PO Box 378, Canberra, 2601, Australia
           \and
           Observatoire de Paris, SYRTE, CNRS, UPMC, GRGS, Paris, France}
\date{}
\abstract
% context heading (optional)
{}
% aims heading (mandatory)
{We propose new estimates of the secular aberration drift, mainly due to the rotation of the Solar System about the Galactic center, based on up-to-date VLBI observations and and improved method of outlier elimination.}
% methods heading (mandatory)
{We fit degree-2 vector spherical harmonics to extragalactic radio source proper motion field derived from geodetic VLBI observations spanning 1979--2013. We pay particular attention to the outlier elimination procedure to remove outliers from (i) radio source coordinate time series and (ii) the proper motion sample.}
% results heading (mandatory)
{We obtain more accurate values of the Solar system acceleration compared to those in our previous paper. The acceleration vector is oriented towards the Galactic center within $\sim$7$^{\circ}$. The component perpendicular to the Galactic plane is statistically insignificant. We show that an insufficient cleaning of the data set can lead to strong variations in the dipole amplitude and orientation, and statistically biased results.}
% conclusions heading (optional), leave it empty if necessary
{}
\titlerunning{}

\maketitle

\section{Introduction}

The accelerated motion of the Solar System in the Universe, mainly due to its rotation about the Galactic center, induces an apparent proper motion of extragalactic objects of a few microseconds of arc per year ($\mu$as/yr) called secular aberration drift, in the direction of the acceleration vector. This effect shows up in the systematic part of the proper motion field.

The detection of the secular aberration drift in the positions of extragalactic radio sources observed by very long baseline interferometry (VLBI) in accordance with the theoretical predictions (see, e.g., Fanselow 1983, Bastian~1995, Eubanks et al.~1995, Sovers et al.~1998, Mignard 2002, Kovalevsky 2003, Kopeikin \& Makarov 2006) was announced recently by Titov et al.~(2011), hereafter TLG11, in the following, and Xu et al.~(2012, 2013). These works provided direct measurements of the Solar System acceleration in the Universe independent from any dynamical model of the Galaxy. TLG11 found an acceleration vector pointing towards the Galactic center ($\alpha_{\rm G}=266.4^{\circ}$, $\delta_{\rm G}=-28.9^{\circ}$) within 10$^{\circ}$ and an amplitude in agreement with predictions based on the Galactic parameters derived with other methods. Using a similar observational data set but an alternative estimation method, Xu et al.~(2012, 2013) obtained an amplitude close to TLG11 but oriented about 18$^{\circ}$ north and 23$^{\circ}$ west from the Galactic center, and thus a significant acceleration of the Solar System perpendicularly to the Galactic plane. The authors raised the possibility of a companion star orbiting the Sun to explain the deviation.

This current study brings out improved estimates of the secular aberration drift, making use of new VLBI observations since TLG11. Section~2 presents some generalities about the Solar System motion and the expected amplitude of the secular aberration drift. Section~3 recalls the basic equations. In Section~4, we study the sensitivity of the solution to the outliers and show how they can dramatically corrupt the least-squares estimates of the dipole components.

\section{The Solar System motion in the Galaxy}

The theoretical effects of the Solar System acceleration on the apparent position of distant bodies are described in many articles (see TLG11 and references therein). Recent estimates of the Galactic parameters based on trigonometric parallaxes of massive star regions (Reid et al.~2009) give distances $R$ to the Galactic center of the order of $8.4\pm0.6$~kpc and a circular rotation speed $V$ of $254\pm16$~km/s (thus a rotation period of $\sim$200~Myr). Consistent values are obtained by other methods, for instance stellar orbit monitoring (see, e.g., Ghez et al. 2008, Gillesen et al. 2009). The acceleration $V^2/R$ deduced from these values equals $7.9\pm1.6$~mm/s/yr. This induces a dipolar proper motion to distant bodies of amplitude $5.4\pm0.7$~$\mu$as/yr towards the Galactic center.

\begin{figure}[htbp]
\begin{center}
\includegraphics[width=8.5cm]{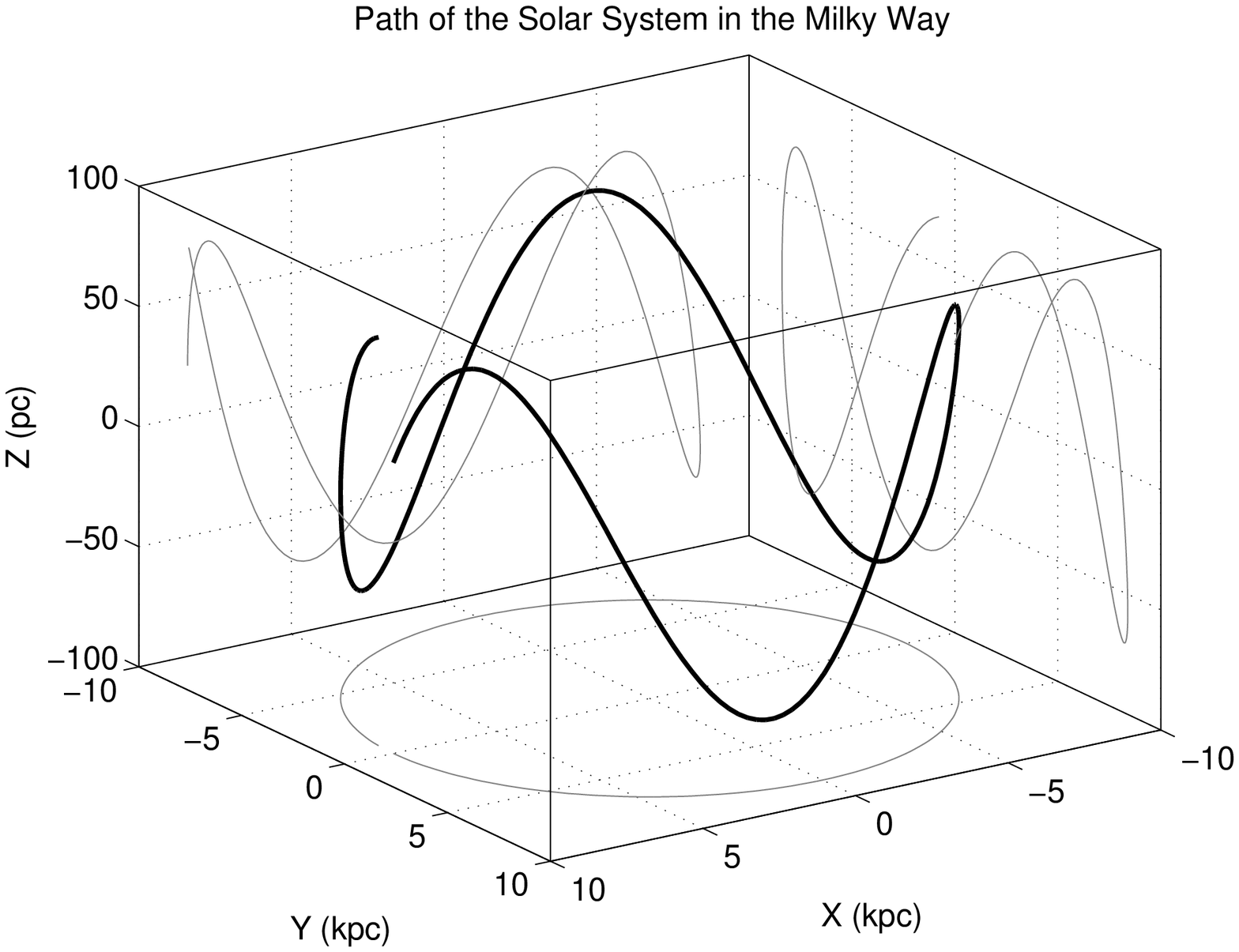}
\includegraphics[width=8.5cm]{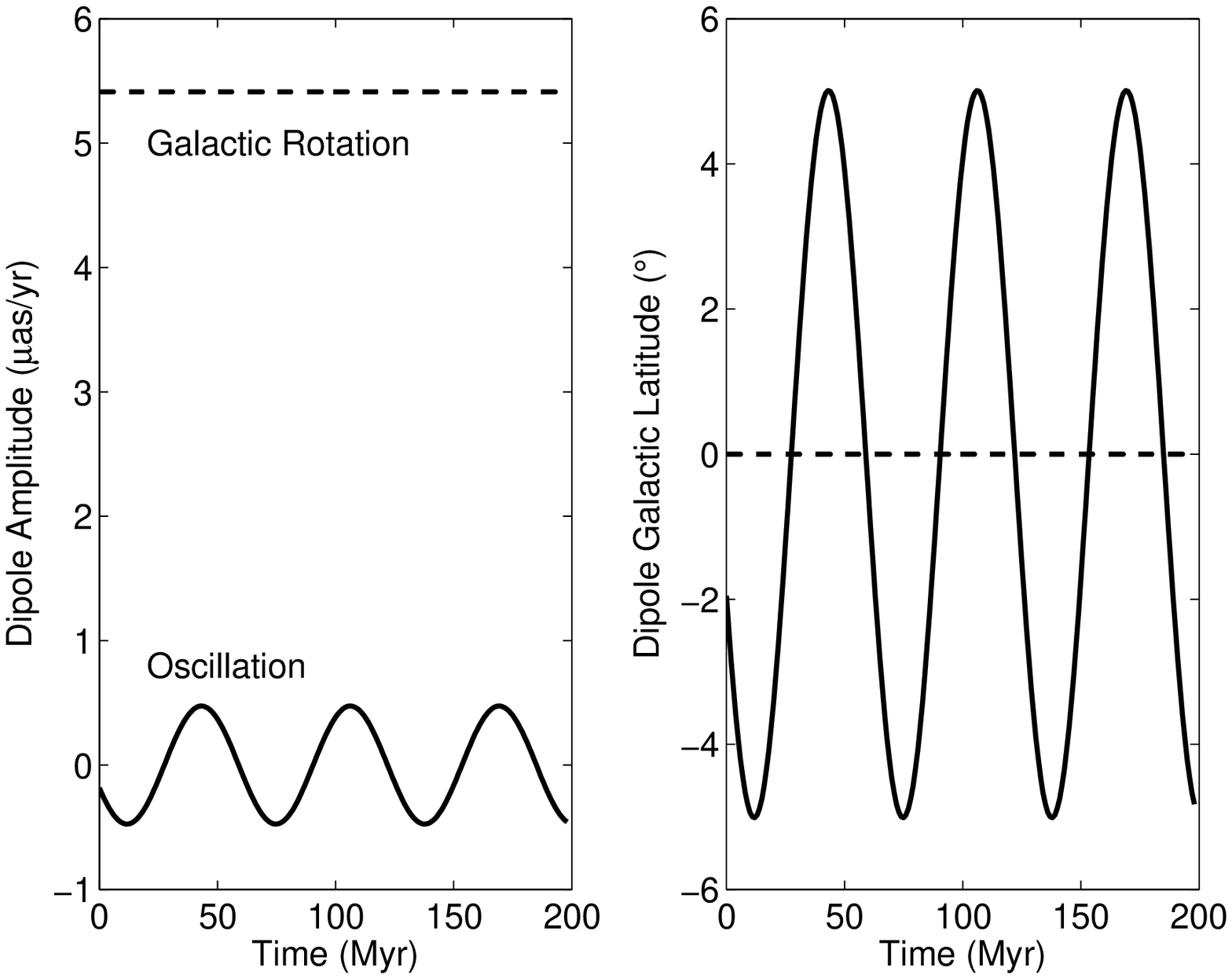}
\end{center}
\caption{({\it Top}) Simulated path of the Solar system over about one revolution about the Galactic center, and contributions to the dipole ({\it Bottom-left}) amplitude and ({\it Bottom-right}) orientation, for averaged oscillation period of 63~Myr and amplitude of 71~pc, and assuming the Solar system crossed the Galactic plane 3~Myr ago.}
\label{fig-simu}
\end{figure}

As the Solar System rotates around the Galactic center, it also oscillates around the Galactic plane with an amplitude of 49--93~pc and a period of 52--74~Myr (Bahcall \& Bahcall 1985). The Solar System passed through the Galactic plane about 3~Myr ago and is now about 26~pc above it (Majaess et al. 2009). It is therefore moving towards the north Galactic pole and slowing down before going back to the south. The magnitude of the aberration on distant body proper motion resulting from this oscillation can reach $\sim$0.5~$\mu$as/yr when the Solar system reaches the maximum distance from the Galactic plane. When this smaller contribution is added to the main contribution from the Galactic rotation, the apparent direction of the dipole can be displaced by about 5$^{\circ}$ from the Galactic center (Fig.~\ref{fig-simu}).

\section{The proper motion field parameters}

We decomposed the systematic part of the proper motion field into a dipole, a rotation, and a quadrupole. Equations~(8)--(9) of TLG11 appear to be vitiated. We corrected these formulae in the present paper (Eqs.~(5)--(6)) and we followed the convention of Mignard \& Klioner~(2012). Corrected TLG11 values are given later in this paper in Table~\ref{tab-best}. For a distant body of equatorial coordinates $(\alpha,\delta)$, the dipole part reads (see, e.g., Mignard \& Morando 1990, Mignard \& Klioner 2012)
\begin{eqnarray} \label{dip}
\Delta\mu_{\alpha}\cos\delta&=&-d_1\sin\alpha+d_2\cos\alpha, \\
\Delta\mu_{\delta}&=&-d_1\cos\alpha\sin\delta-d_2\sin\alpha\sin\delta+d_3\cos\delta,
\end{eqnarray}
where the $d_i$ are the components of the acceleration vector in unit of the proper motion. In addition to the aberration distortion, there may also be a small global rotation that can be described by the toroidal harmonics of degree~1:
\begin{eqnarray} \label{rot}
\Delta\mu_{\alpha}\cos\delta&=&r_1\cos\alpha\sin\delta+r_2\sin\alpha\sin\delta-r_3\cos\delta, \\
\Delta\mu_{\delta}&=&-r_1\sin\alpha+r_2\cos\alpha.
\end{eqnarray}
The quadrupolar anisotropy of the proper motion field is given by the development of the degree 2 vector spherical harmonics of electric ($E$) and magnetic ($M$) types:
\begin{eqnarray} \label{quad}
\Delta\mu_{\alpha}\cos\delta&=&a_{2,0}^M\sin 2\delta \nonumber \\
                            &+&\sin\delta\left(a_{2,1}^{E,\rm Re}\sin\alpha+a_{2,1}^{E,\rm Im}\cos\alpha\right) \nonumber \\
                            &-&\cos 2\delta\left(a_{2,1}^{M,\rm Re}\cos\alpha-a_{2,1}^{M,\rm Im}\sin\alpha\right) \nonumber \\
                            &-&2\cos\delta\left(a_{2,2}^{E,\rm Re}\sin 2\alpha+a_{2,2}^{E,\rm Im}\cos 2\alpha\right) \nonumber \\
                            &-&\sin 2\delta\left(a_{2,2}^{M,\rm Re}\cos 2\alpha-a_{2,2}^{M,\rm Im}\sin 2\alpha\right), \\
\Delta\mu_{\delta}&=&a_{2,0}^E\sin 2\delta \nonumber \\
                  &-&\cos 2\delta\left(a_{2,1}^{E,\rm Re}\cos\alpha-a_{2,1}^{E,\rm Im}\sin\alpha\right) \nonumber \\
                  &-&\sin\delta\left(a_{2,1}^{M,\rm Re}\sin\alpha+a_{2,1}^{M,\rm Im}\cos\alpha\right) \nonumber \\
                  &-&\sin 2\delta\left(a_{2,2}^{E,\rm Re}\cos 2\alpha-a_{2,2}^{E,\rm Im}\sin 2\alpha\right) \nonumber \\
                  &+&2\cos\delta\left(a_{2,2}^{M,\rm Re}\sin 2\alpha+a_{2,2}^{M,\rm Im}\cos 2\alpha\right).
\end{eqnarray}

In the following, we consider two solutions: the DR solution only estimates dipole and rotation parameters and the DRQ solution estimates the 16 parameters relevant to the dipole, the rotation, and the quadrupole.

\begin{figure}[htbp]
\begin{center}
\includegraphics[width=8.5cm]{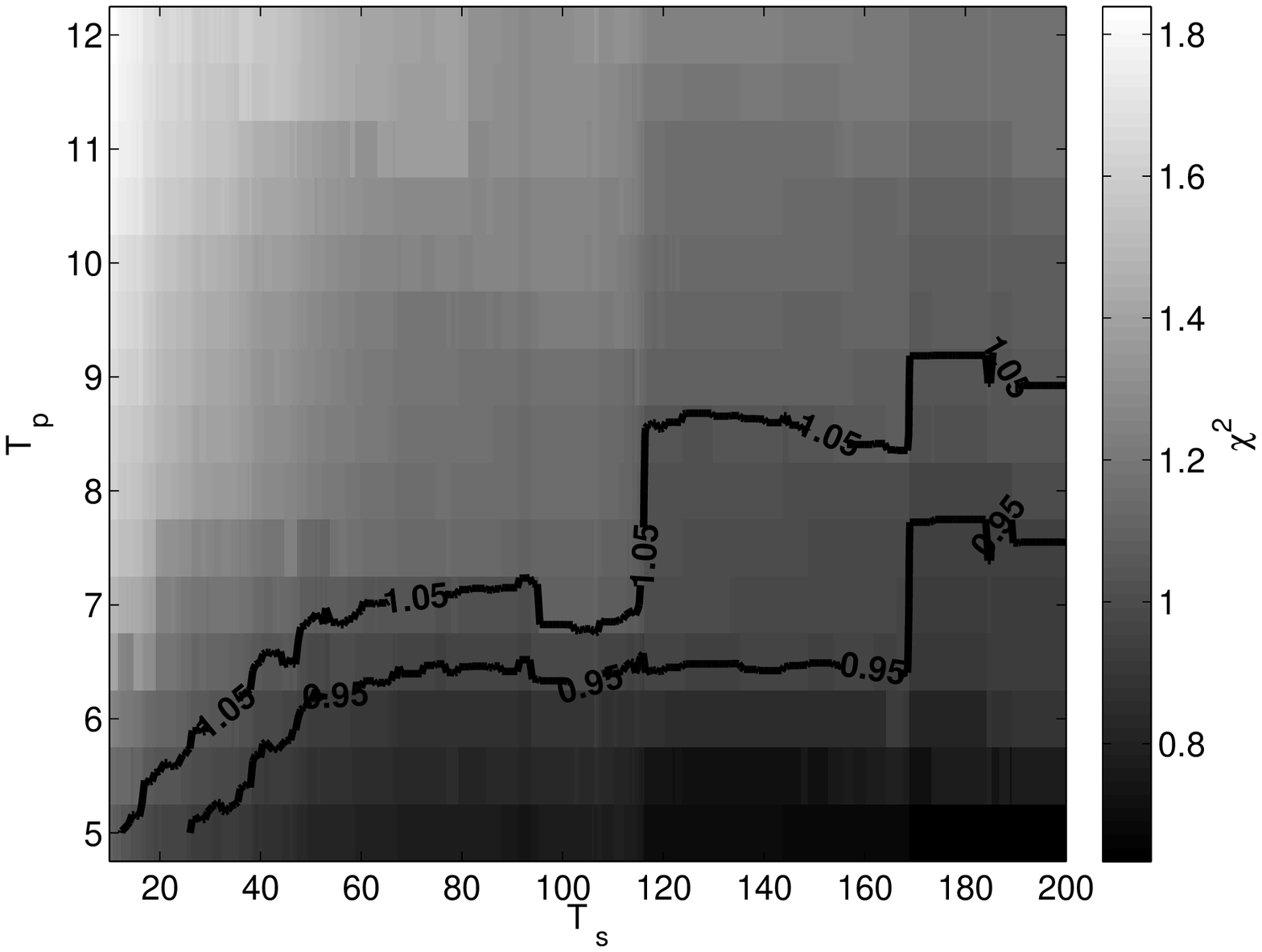}
\includegraphics[width=8.5cm]{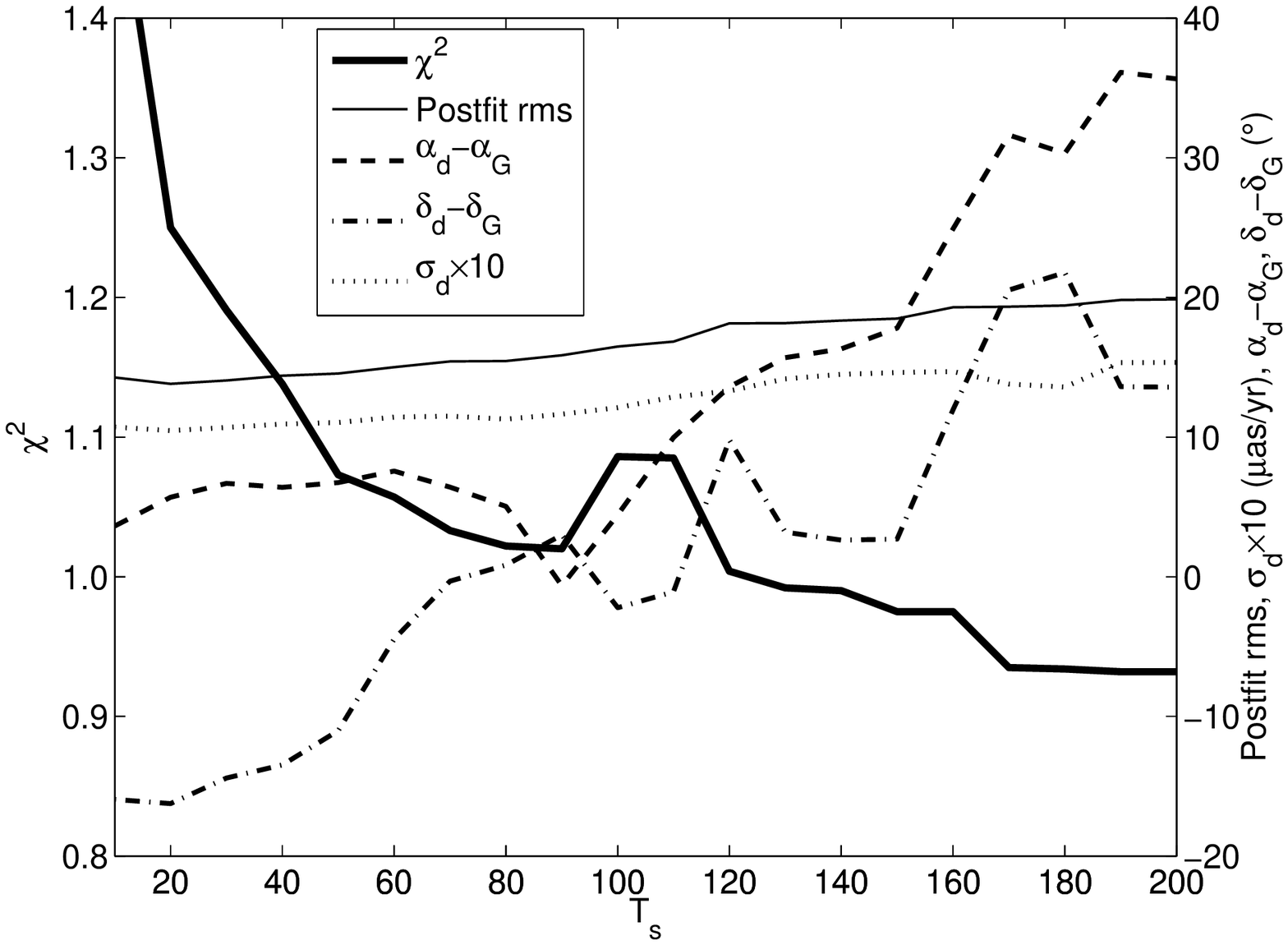}
\end{center}
\caption{({\it Top}) Reduced $\chi^2$ vs. $T_{\rm s}$ and $T_{\rm p}$. The contour indicates the region wherein $0.95<\chi^2<1.05$. ({\it Bottom}) Results of the adjustment vs. $T_{\rm s}$ obtained with $T_{\rm p}=7$. $\sigma_{\rm d}$ is the formal error on the dipole amplitude.}
\label{fig-sens}
\end{figure}

\begin{table}
\begin{center}
\begin{tabular}{lrrr}
\hline
\hline
\noalign{\smallskip}
& DR & DRQ & DRQ11 \\
\noalign{\smallskip}
\hline
\noalign{\smallskip}
No. sources & 429 & 426 & 555 \\
\noalign{\smallskip}
\hline
\noalign{\smallskip}
Dipole & & & \\
\noalign{\smallskip}
$d_1$               & $-0.4 \pm 0.7$ & $ 0.7 \pm 0.8$ & $ 0.7 \pm 0.9$ \\
$d_2$               & $-5.7 \pm 0.8$ & $-6.2 \pm 0.9$ & $-6.2 \pm 1.0$ \\
$d_3$               & $-2.8 \pm 0.9$ & $-3.3 \pm 1.0$ & $-3.3 \pm 1.0$ \\
\noalign{\smallskip}
Amplitude                   & $ 6.4 \pm 1.1$ & $ 7.1 \pm 1.3$ & $ 7.1 \pm 1.4$ \\
Direction in $\alpha$ ($^{\circ}$) & $266\pm 7$ & $277\pm 7$ & $277 \pm 9$ \\
Direction in $\delta$ ($^{\circ}$) & $-26\pm 7$ & $-28\pm 7$ & $-28 \pm 8$ \\
\noalign{\smallskip}
\hline
\noalign{\smallskip}
Rotation & & & \\
\noalign{\smallskip}
$r_1$                & $-1.1 \pm 0.9$ & $-0.5 \pm 1.2$ & $-2.4 \pm 0.8$ \\
$r_2$                & $ 1.4 \pm 0.8$ & $ 1.8 \pm 1.3$ & $ 0.4 \pm 1.0$ \\
$r_3$                & $ 0.7 \pm 0.6$ & $ 0.8 \pm 1.4$ & $ 0.8 \pm 0.7$ \\
\noalign{\smallskip}
\hline
\noalign{\smallskip}
Quadrupole & & & \\
\noalign{\smallskip}
$a_{2,0}^E$ &            & $ 1.3 \pm 1.0$ & $ 2.8 \pm 1.2$ \\
$a_{2,0}^M$ &            & $ 0.3 \pm 0.8$ & $-0.7 \pm 0.9$ \\
$a_{2,1}^{E,{\rm Re}}$ & & $ 1.7 \pm 1.0$ & $ 4.1 \pm 1.3$ \\
$a_{2,1}^{E,{\rm Im}}$ & & $ 0.9 \pm 1.1$ & $ 1.7 \pm 1.2$ \\
$a_{2,1}^{M,{\rm Re}}$ & & $-1.8 \pm 1.0$ & $-0.6 \pm 1.1$ \\
$a_{2,1}^{M,{\rm Im}}$ & & $ 2.3 \pm 0.9$ & $ 2.0 \pm 1.1$ \\
$a_{2,2}^{E,{\rm Re}}$ & & $ 0.5 \pm 0.5$ & $ 0.8 \pm 0.5$ \\
$a_{2,2}^{E,{\rm Im}}$ & & $-1.0 \pm 0.4$ & $ 0.3 \pm 0.5$ \\
$a_{2,2}^{M,{\rm Re}}$ & & $ 2.3 \pm 0.6$ & $ 1.2 \pm 0.6$ \\
$a_{2,2}^{M,{\rm Im}}$ & & $-1.3 \pm 0.6$ & $-0.3 \pm 0.6$ \\
\noalign{\smallskip}
\hline
\noalign{\smallskip}
Postfit rms & 15.86 & 15.64 & 21.56 \\
$\chi^2$ & 1.02 & 1.00 & 1.91 \\
\noalign{\smallskip}
\hline
\end{tabular}
\end{center}
\caption{Estimated parameters ($\mu$as/yr) of the proper motion field. The rightest column (DRQ11) reports the TLG11 Table~3 values obtained using quadrupole Eqs.~(5)--(6). Uncertainties are 1$\sigma$.}
\label{tab-best}
\end{table}

\begin{figure}[htbp]
\begin{center}
\includegraphics[width=8.5cm]{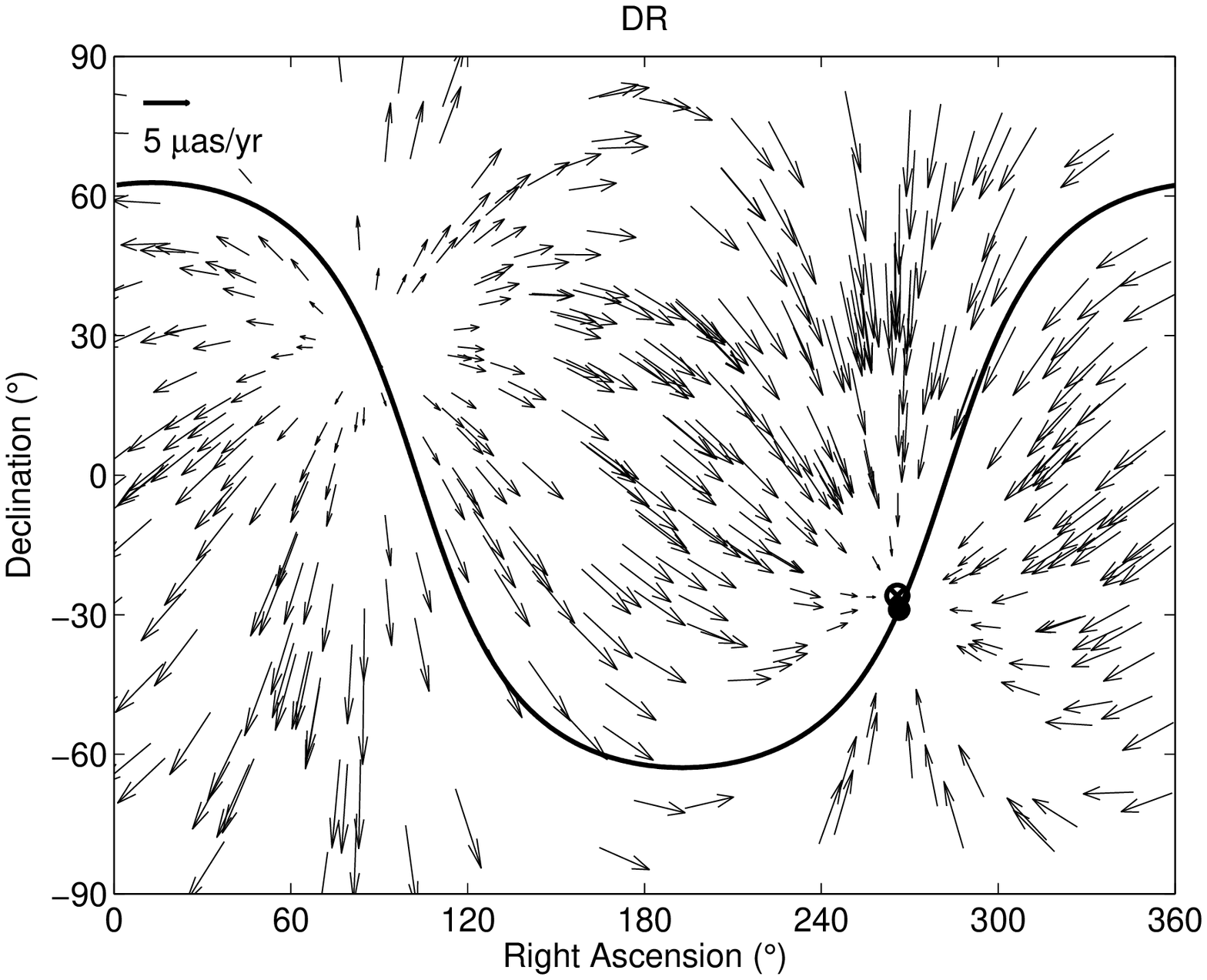}
\includegraphics[width=8.5cm]{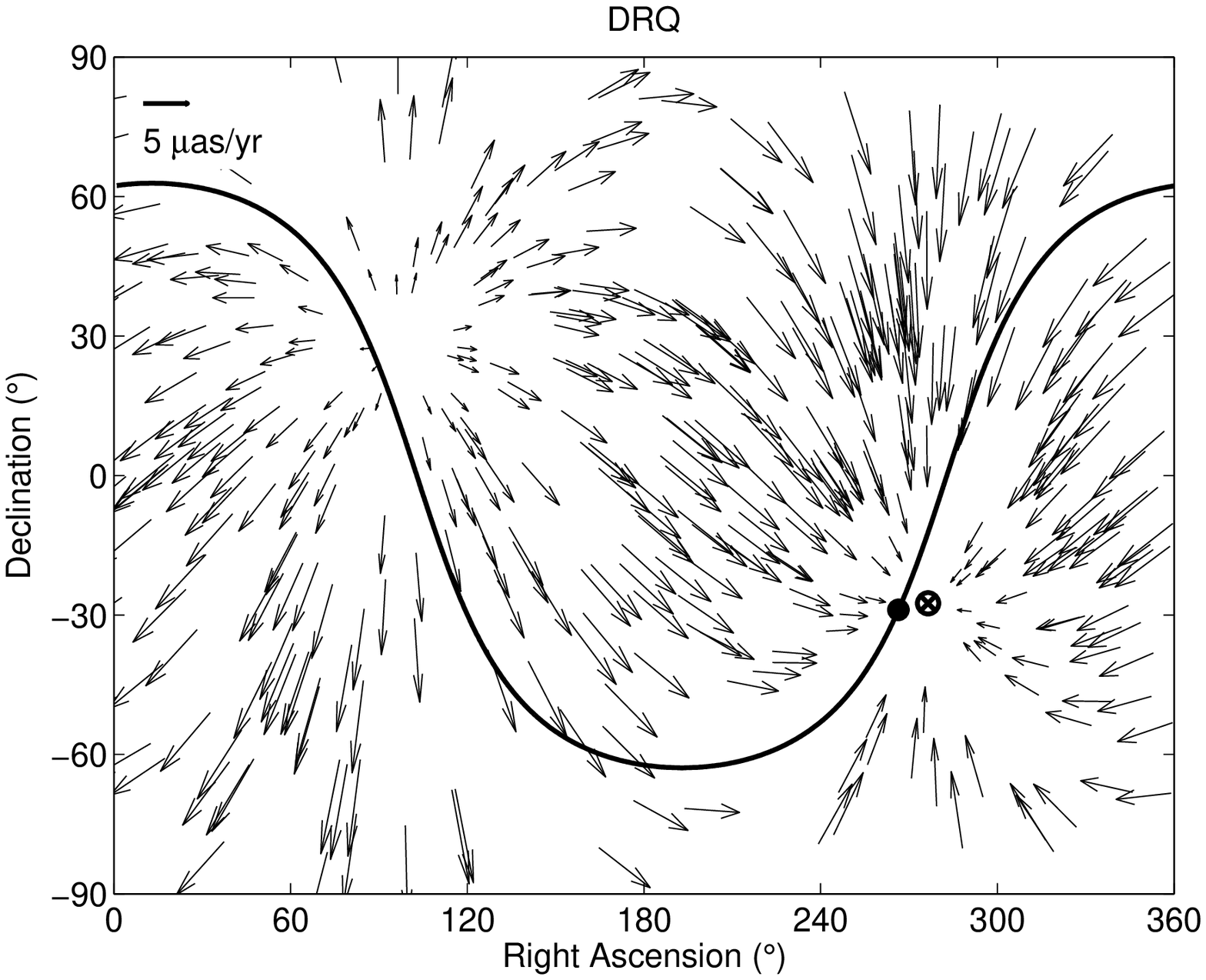}
\end{center}
\caption{Dipole pattern obtained from the ({\it Top}) DR and ({\it Bottom}) DRQ solutions. The solid line indicates the Galactic plane. The dashed line represents the ecliptic. The black disk and the circled cross indicate the Galactic center and the direction of the dipole, respectively.}
\label{fig-est}
\end{figure}

\section{Data analysis and results}

The VLBI data processing made use of the analysis options extensively described in TLG11. We extend the session list up to the end of February 2013. Due to the major 2011 Tohoku earthquake, several Japanese telescopes in the Tokyo area were displaced by several tens of centimeters. Consequently, these stations were removed from the no-net rotation and translation constraints, as previously done for the Fairbanks and Concepci\'on antennas. As in TLG11, we applied a loose constraint of no-net rotation on the radio source coordinates, with the exception of the 39 sources showing significant nonlinear positional variations due to large scale variations in their structure (e.g., 3C84, 3C273B, 3C279, 3C345, 3C454.3, 4C39.25) pointed out in Fey et al.~(2010). In TLG11, we emphasized that the constraint should be loose enough to allow the reference frame axes to deform slowly.

The analysis returned coordinate time series for 3635 extragalactic radio sources, of which 707 were observed in more than one session. In each time series, data points obtained from less then three delays were removed. Data points whose distance to the mean is higher than a certain threshold $T_{\rm s}$ times the uncertainty were also removed. This elimination is repeated until the $\chi^2$ is reasonably close to unity. The removed outliers are generally associated with unreliable networks or corrupted data.

Proper motions were then computed by the least-squares fit to time series longer than 10 sessions weighted by the inverse of the squared errors. (The influence of the minimum number of sessions was checked afterwards and found to be negligible.) Note that the VLBI analysis software package used here allows direct computation of proper motions from VLBI delays. Proper motions obtained by this method appear to be very different from those obtained by fitting to time series. The difference may come from the fact that the software package does not remove bad observations and gets therefore biased estimates. Such discrepancies should be investigated by the community in the future.

The next step consisted of fitting the proper motion field parameters. Fitting the dipole and rotation parameters to the data with $T_{\rm s}$ in a range of 10 to 100 gives a dipole of amplitude $\sim$6~$\mu$as/yr oriented towards $\alpha\sim280\pm15^{\circ}$ and $\delta\sim-35\pm15^{\circ}$, which is consistent with the values expected from the theory. However, proper motions adjusted in the previous step can be unreliable due to two major reasons. Firstly, strong, frequently observed radio source have very small formal errors. However, some of them show very large nonlinear positional variations and/or significant apparent motion due to relativistic jets and intrinsic radio source structure (Charlot~1990, Fey et al.~2004). Fortunately, these astrometrically unstable radio sources are well known in advance and could be removed from the sample. Secondly, some radio sources with a small number of observations can exhibit proper motions extremely large compared to the global rms. These radio sources have the same impact on the solution as the frequently observed ones and should be removed from the sample. The problem is that these radio sources are not known in advance. They can only be identified after a preliminary solution by inspecting the proper motion sample. To identify and remove these spurious proper motions, the systematics were first estimated and removed, and then the source with residual velocities larger than a threshold $T_{\rm p}$ times the residual rms were eliminated from the sample. The process was iterated a few times to convergence.

To illustrate the usefullness of such an elimination algorithm, we took $T_{\rm s}=90$ and estimated the dipole parameters through the obtained velocity field. We obtained 6.3~$\mu$as/yr, $\alpha=275^{\circ}$, $\delta=-30^{\circ}$. Then, we introduced a fake source located at the Vernal point with a spurious proper motion of magnitude 10~mas/yr $\pm$ 50~$\mu$as/yr both in right ascension and declination. The fit gave 3.5~$\mu$as/yr, $\alpha=304^{\circ}$, $\delta=2^{\circ}$.

We tried several values of the thresholds $T_{\rm s}$ and $T_{\rm p}$. Figure~\ref{fig-sens} displays the results of the different adjustments with a contour delimiting the region where the $\chi^2$ is close to~1. This occurs for $T_{\rm p}$ preferably close to~7. We took this value to check the sensitivy of the fit to $T_{\rm s}$ (bottom panel of Fig.~\ref{fig-sens}). The region for which $\chi^2$ nears~1 with the lower postfit rms is narrow and located around $T_{\rm s}=90$. In this region, the dipole gets oriented towards the Galactic center. Lowering $T_{\rm s}$ leads to a lower formal error on the dipole parameters but also to a $\chi^2$ significantly far from~1. In contrast, larger values of $T_{\rm s}$ (equivalent to imposing no outlier elimination) produce large offsets to the Galactic center and the departure of the $\chi^2$ from unity.

\begin{table}
\begin{center}
\begin{tabular}{lrrrrr}
\hline
\hline
\noalign{\smallskip}
& $d_1$ & $d_2$ & $d_3$ & $r_1$ & $r_2$ \\
\noalign{\smallskip}
\hline
\noalign{\smallskip}
$d_2$ & $  0.07$& & & & \\
$d_3$ & $  0.00$& $-0.01$&         & & \\
$r_1$ & $ -0.02$& $-0.41$& $  0.12$&         & \\
$r_2$ & $  0.48$& $ 0.03$& $  0.02$& $ -0.03$&     \\
$r_3$ & $ -0.11$& $ 0.02$& $  0.00$& $ -0.06$& $  0.01$ \\
\noalign{\smallskip}
\hline
\end{tabular}
\end{center}
\caption{Correlations between parameters in the DR solution.}
\label{tab-cordr}
\end{table}

\begin{table*}
\begin{center}
\begin{tabular}{lrrrrrrrrrrrrrrr}
\hline
\hline
\noalign{\smallskip}
& $d_1$ & $d_2$ & $d_3$ & $r_1$ & $r_2$ & $r_3$ & $a_{2,0}^E$ & $a_{2,0}^M$ & $a_{2,1}^{E,{\rm Re}}$ & $a_{2,1}^{E,{\rm Im}}$ & $a_{2,1}^{M,{\rm Re}}$ & $a_{2,1}^{M,{\rm Im}}$ & $a_{2,2}^{E,{\rm Re}}$ & $a_{2,2}^{E,{\rm Im}}$ & $a_{2,2}^{M,{\rm Re}}$ \\
\noalign{\smallskip}
\hline
\noalign{\smallskip}
$d_2$                   & $  0.06$& & & & & & & & & & & & & & \\
$d_3$                   & $  0.00$&$  0.00$& & & & & & & & & & & & & \\
$r_1$                   & $ -0.01$&$ -0.43$&$  0.09$& & & & & & & & & &\\
$r_2$                   & $  0.46$&$  0.01$&$  0.01$&$ -0.02$& & & & & & & & \\
$r_3$                   & $ -0.03$&$  0.05$&$  0.00$&$ -0.04$&$  0.03$& & & & & & & \\
$a_{2,0}^E$             & $ -0.01$&$  0.08$&$ -0.40$&$  0.03$&$  0.02$&$  0.00$& & & & & &\\
$a_{2,0}^M$             & $  0.03$&$  0.10$&$  0.00$&$  0.02$&$ -0.04$&$  0.39$&$  0.00$& & & & & & & \\
$a_{2,1}^{E,{\rm Re}}$  & $  0.36$&$  0.01$&$ -0.01$&$ -0.03$&$ -0.08$&$  0.05$&$ -0.03$&$ -0.07$& & & & & & & \\
$a_{2,1}^{E,{\rm Im}}$  & $ -0.02$&$ -0.33$&$ -0.05$&$ -0.02$&$ -0.03$&$ -0.05$&$ -0.12$&$  0.03$&$ -0.04$& & & & & & \\
$a_{2,1}^{M,{\rm Re}}$  & $ -0.02$&$  0.34$&$  0.00$&$ -0.34$&$  0.02$&$ -0.03$&$  0.07$&$ -0.03$&$  0.03$&$ -0.18$& & & & & \\
$a_{2,1}^{M,{\rm Im}}$  & $  0.33$&$ -0.03$&$ -0.01$&$ -0.02$&$  0.40$&$  0.05$&$ -0.02$&$ -0.15$&$  0.22$&$ -0.04$&$  0.07$& & & & \\
$a_{2,2}^{E,{\rm Re}}$  & $ -0.01$&$  0.08$&$  0.00$&$  0.06$&$ -0.05$&$ -0.13$&$  0.01$&$  0.02$&$ -0.09$&$  0.06$&$ -0.07$&$ -0.03$& & & \\
$a_{2,2}^{E,{\rm Im}}$  & $  0.02$&$  0.04$&$  0.01$&$ -0.01$&$  0.00$&$  0.11$&$ -0.01$&$  0.01$&$  0.01$&$  0.01$&$ -0.06$&$ -0.02$&$  0.00$& & \\
$a_{2,2}^{M,{\rm Re}}$  & $  0.02$&$ -0.04$&$  0.00$&$ -0.03$&$ -0.11$&$ -0.03$&$  0.03$&$ -0.10$&$  0.13$&$  0.04$&$ -0.03$&$  0.01$&$ -0.03$&$ -0.32$& \\
$a_{2,2}^{M,{\rm Im}}$  & $ -0.04$&$ -0.03$&$ -0.04$&$ -0.05$&$  0.00$&$ -0.02$&$ -0.03$&$ -0.06$&$  0.05$&$ -0.04$&$  0.05$&$  0.06$&$  0.28$&$  0.02$&$ -0.03$ \\
\noalign{\smallskip}
\hline
\end{tabular}
\end{center}
\caption{Correlations between parameters in the DRQ solution.}
\label{tab-cordrq}
\end{table*}

Table~\ref{tab-best} reports the dipole, rotation, and quadrupole parameters obtained using $T_{\rm s}=90$ and $T_{\rm p}=7$. The rightest column of this table also shows results of Table~3 of TLG11 obtained when using Eqs.~(5)--(6) of the present paper. With respect to TLG11, the dipole standard error has improved by about 20\% and the postfit rms was reduced by 28\%. No statistically significant rotation and quadrupole harmonics were found. Tables~\ref{tab-cordr} and \ref{tab-cordrq} display the correlations between the various parameters which are larger than 0.4 between $d_1$ and $r_2$, $d_2$ and $r_1$, and $d_3$ and $a_{2,0}^E$. The pattern obtained for the dipole is plotted in Fig.~\ref{fig-est}.

The total acceleration of the Solar System barycentre from the DR solution is ($9.3$,~$0.4$,~$0.3$)~$\pm$~(1.1,~1.1,~1.3)~mm/s/yr in the Galactic reference frame. The centripetal acceleration is $9.3\pm1.1$~mm/s/yr. Assuming $R=8.4$~kpc, it is equivalent to a circular rotation speed in the Galactic plane of $282\pm32$~km/s. The acceleration exhibits a non statistically significant component perpendicular to the Galactic plane. For the DRQ solution, the acceleration amounts to ($10.1$,~$1.0$,~$-1.3$)~$\pm$~(1.2,~1.3,~1.4)~mm/s/yr in the Galactic reference frame, equivalent to a rotation speed of $303\pm34$~km/s. The vertical component is also statistically unsignificant.

\section{Conclusion}

This study showed that our previous determination of the secular aberration drift (Titov et al.~2011) is robust after the addition of new VLBI data. The results are consistent with predictions from Galactic models. The quasar proper motion field exhibits a dipole component oriented towards the Galactic center within $\sim$7$^{\circ}$. However, the quadrupole component remains statistically insignificant.

A key point of our computation was the elimination of outlier data points and proper motions. We showed that this step must be considered with great care: if bad data are not properly identified and eliminated, they are likely to perturb significantly the estimates and lead to statistically biased results.

\begin{acknowledgements}
The authors thank Drs. Laura Stanford and Craig Harrison of Geoscience Australia for proof-reading the manuscript and valuable comments. OT has pleasure in acknowledging the financial support from the Paris Observatory which made possible a one-month stay in Paris. This paper has been published with permission of the Chief Executive Officer of Geoscience Australia.
\end{acknowledgements}

\end{document}